\documentclass[9pt, technote, compsoc]{IEEEtran}

\usepackage{fancyhdr}
\usepackage{graphicx}
\usepackage{subfig}
\usepackage{epstopdf}
\usepackage{pifont}
\usepackage{epsfig}
\usepackage{setspace}
\usepackage{amsmath}
\usepackage{multirow}
\usepackage{floatrow}
\usepackage{array}
\usepackage{comment}

\ifCLASSOPTIONcompsoc
  \usepackage[nocompress]{cite}
\else
  \usepackage{cite}
\fi

\usepackage{array}

\ifCLASSINFOpdf
\else
\fi

\begin{document}

\title{SimpleSSD: Modeling Solid State Drives for Holistic System Simulation}
%Holistic System Simulations with A Simplified Solid State Drive Model}
\author{Myoungsoo Jung$^{1}$, Jie Zhang$^{1}$, Ahmed Abulila$^{2}$, Miryeong Kwon$^{1}$, Narges Shahidi$^{3}$, John Shalf$^{4}$, \\
Nam Sung Kim$^{2}$ and Mahmut Kandemir$^{3}$\\
{\emph{$^{1}$Computer Architecture and Memory Systems Lab, Yonsei University}} \\
{\emph{$^{2}$University of Illinois Urbana-Champaign,}}\\
{\emph{$^{3}$Pennsylvania State University,}} \\ 
{\emph{$^{4}$Lawrence Berkeley National Laboratory}} \\
}

%\markboth{COMPUTER ARCHITECTURE AND MEMORY SYSTEMS LABORATORY (CAMEL)}%
%{Shell \MakeLowercase{\textit{et al.}}: Bare Advanced Demo of IEEEtran.cls for Journals}

\IEEEtitleabstractindextext{%
\begin{abstract}
Existing solid state drive (SSD) simulators unfortunately lack hardware and/or software architecture models. Consequently, they are far from capturing the critical features of contemporary SSD devices.
More importantly, while the performance of modern systems that adopt SSDs can vary based on their numerous internal design parameters and storage-level configurations, a full system simulation with traditional SSD models often requires unreasonably long runtimes and excessive computational resources. 
In this work, we propose \texttt{SimpleSSD}\footnotemark[1], a high-fidelity simulator that models all detailed characteristics of hardware and software, while simplifying the nondescript features of storage internals. In contrast to existing SSD simulators, \texttt{SimpleSSD} can easily be integrated into publicly-available full system simulators. In addition, it can accommodate a complete storage stack and evaluate the performance of SSDs along with diverse memory technologies and microarchitectures. Thus, it facilitates simulations that explore the full design space at different levels of system abstraction.  
\end{abstract}
}

\footnotetext[1]{This paper has been accepted at IEEE Computer Architecture Letters (CAL), 2017. This material is presented to ensure timely dissemination of scholarly and technical work. Please refer and cite the IEEE work of this paper \cite{simplessd}}

\maketitle

\IEEEdisplaynontitleabstractindextext
\IEEEpeerreviewmaketitle

\section{Introduction}
\label{sec:intro}
In the past decade, solid state disks (SSDs) have reshaped modern memory hierarchy by replacing conventional spinning disks and/or blurring the boundary between main memory and storage systems. Thanks to their high performance and low power consumption characteristics, SSDs have already become the dominant storage type in diverse computing domains, ranging from embedded to general-purpose and high-performance computing systems.
%The key to success in these scenarios of SSD-based I/O acceleration is to employ the right SSD and place it in right place across the modern memory hierarchy (at the right time).  
This in turn has led to a wide spectrum of research, including the comprehensive exploration of the full design space, storage stack optimization, and architecture renovation at various layers of memory and storage subsystems.% \cite{huang2015unified,lee2016application}. 

While simulations are indispensable for system designers and computer architects, very few SSD simulators have been released to the public domain \cite{prabhakaran2009ssd, flashsim, jung2012nandflashsim, hu2011performance}. Further, these simulators have constraints that prevent them from filling the needs of design space exploration for emerging memory and storage subsystems.  
First, all existing SSD simulators lack system-level simulation capability, and integrating these simulators with publicly-available full-system simulators is a non-trivial task. While the execution of a CPU instruction only takes a few cycles in a simulation, a storage access requires tens of millions (even billions) of cycles for its service. Similarly, a file access in an accurate SSD simulation model can exhibit a long execution time because it needs to go through the SSD's intricate software stack and hardware architecture. Traditional SSD simulators cannot fully account for the important functionalities of the underlying firmware and model the underlying hardware in detail. Thus, they are far from capturing the critical features of contemporary high-performance SSD architectures.

In this work, we propose \texttt{SimpleSSD}, a high-fidelity simulator that models all of the detailed characteristics of hardware and software while simplifying the nondescript features of storage internals such as multi-cycle operations to address a target page on a flash interface. The proposed hardware and software simplifications allow \texttt{SimpleSSD} to accommodate a complete storage stack. Thus, system designers and computer architects can evaluate the SSDs performance along with diverse memory technologies and can explore the full design space of an SSD architecture. Moreover, \texttt{SimpleSSD} can easily be integrated with publicly-available full-system simulators and can capture relevant CPU performance characteristics impacted by different storage types employed by the system. %different storage that the system employed. 
As a case study, we integrated \texttt{SimpleSSD} with the popular full-system simulator, gem5 \cite{binkert2011gem5}, and evaluated its system-level performance from various aspects. 
Note that traditional SSD simulators \cite{prabhakaran2009ssd, flashsim, hu2011performance} capture only storage-related metrics such as bandwidth and latency by replaying block-level I/O traces; this ignores system-level interaction between the host-side CPU and storage subsystems. %information and make storage subsystems disconnected from the host simulations, 
In contrast, the proposed \texttt{SimpleSSD} can report detailed information from low-level memory to each firmware module in order to determine the host-side CPU performance while executing entire applications. The SimpleSSD source code can be freely downloaded from the following website: http://simplessd.camelab.org.
%\emph{All \texttt{SimpleSSD} simulation sources and tutorials will be freely downloaded.}

\begin{figure}
%\vspace{-10pt}
\centering{}\includegraphics[width=1\columnwidth]{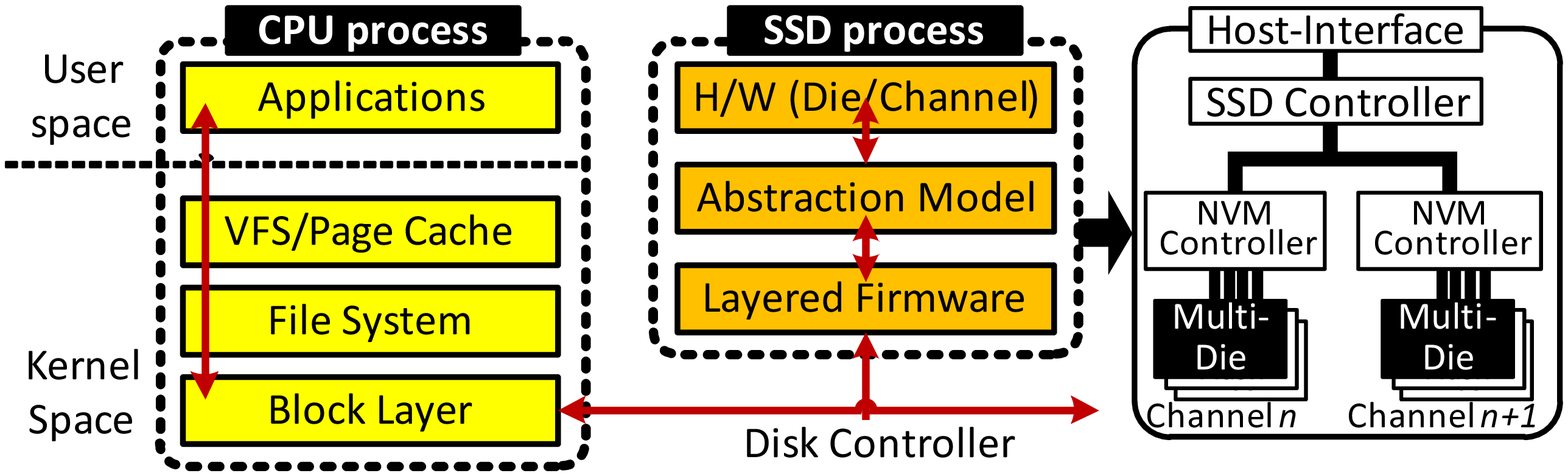}
\caption{Overview of \texttt{SimpleSSD}.\label{fig:ssd-gem5}} 
\end{figure}

\section{SSD-Enabled System Simulation Overview}

Figure \ref{fig:ssd-gem5} shows an overview of a holistic system simulation with the proposed \texttt{SimpleSSD}. 
Application(s) simulated on the host can place an I/O request through a virtual file system (VFS) and native file system. The VFS buffers small-sized requests through a page cache, whereas the native file system manages the data accesses and system memory. The request then arrives at a block layer that reorders and combines multiple requests into a specific order. This CPU processing part can communicate with the layered firmware of \texttt{SimpleSSD} via a disk controller. Then, the layered firmware simulates the SSD process part by interacting with an abstraction model, which simulates the given SSD hardware architecture including multiple flash dies, module interfaces, and channels. Although \texttt{SimpleSSD} leveraged gem5 running in full-system mode to simulate such CPU processing in this study, it can easily be integrated into other full-system simulators such as MARSSx86 \cite{patel2011marss}.

\begin{figure*}
%\vspace{-10pt}
\centering{}\includegraphics[width=1\columnwidth]{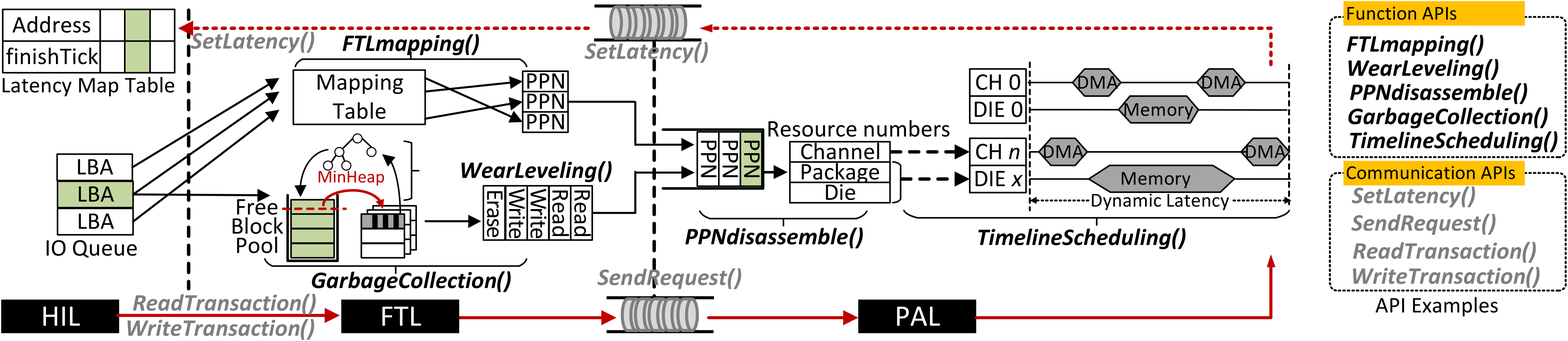}
\caption{High-level view of \texttt{SimpleSSD}.\label{fig:HIL}} 
\end{figure*}

\noindent \textbf{Layered firmware.} %While the performances of SSDs significantly vary based on how internal firmware manages the underlying flash, the existing simulation models only support a fixed address mapping mechanism. 
One of the main challenges of simulating an SSD is supporting diverse flash firmware versions, which greatly influences the target storage performance. We model a flexible \emph{flash translation layer} (FTL) whose address translation mechanism can simply be reconfigured based on different associativity granularities defined by system architects. We also decouple I/O scheduling and page allocation mechanisms from the FTL so that new scheduling proposals that are aware of SSD-internal parallelism can be embedded without changing the FTL. Although we do not cover all types of potential FTLs, the implemented reconfigurable mapping algorithm can capture/support diverse operational characteristics of a block-level mapping FTL, a fully-associative FTL, and various hybrid mapping schemes that employ different levels of block and page mapping tables in their address translations. In addition, our simplified but reconfigurable layered firmware also offers diverse research opportunities where system and computer architects can simply modify some performance-critical components such as garbage collection and wear-leveling algorithms with different mapping mechanisms. 

% \cite{alfke2009xilinx} 
\noindent \textbf{Hardware abstraction.} The performance characteristics of the underlying hardware vary based on i) the intrinsics of latency  of individual flash characteristics and ii) their different levels of parallelism. A cycle-level simulation for each component can accurately evaluate all SSD internals. However, full-system simulations with an SSD at the cycle level require an unreasonably long runtime and excessive resources. In this work, we abstracted both flash-level and subsystem-level hardware characteristics. We implemented an FPGA-based memory controller built on Xilinx Spartan-6 and then used this to characterized different memory technologies. Based on the extracted characteristics, we first design a die-level latency model by simplifying the flash transactions. Specifically, we examined all flash transactions specified by the open NAND flash interface (ONFi 3.x \cite{workgroup2011open}) and classified various timing components of the corresponding protocol into a few transaction activities. With this simplified latency model, the proposed \texttt{SimpleSSD} simulates varying numbers of flash chips over many interconnection buses by modeling the executions across different hardware resources and resource contentions. Even though this simplified model cannot account for all of the characteristics from the flash at a cycle level, it can capture the close interactions among the designs of the firmware, controller, and architecture by being aware of flash latency intrinsics and internal parallelism.

%\noindent $\bullet$ \emph{SSD models for full system simulation.} Thanks to our new hardware abstraction and simplified firmware model, the proposed \texttt{SimpleSSD} can be integrated into a full system simulator and be simulated with the execution of diverse kernel or user applications.  As an example of user scenarios, in this work, we integrate \texttt{SimpleSSD} with a popular full system simulator, gem5 \cite{binkert2011gem5}. Specifically, we demonstrate an active SSD by executing entire applications of an industry-standardized CPU benchmark (SPEC2006 \cite{}) on system call emulation mode. Unlike the existing simulators, in this active SSD use case, \texttt{SimpleSSD} can report data-processing bandwidth, DRAM operations, and the execution time of each application by considering various flash technologies and architectures. On the other hand, we also evaluate several file system benchmarks with the executions of their images on full system mode, and measure the system-level performance such as instruction per second (IPC) by employing different SSD configurations. Note that while the existing SSD simulators capture only storage performance such as bandwidth or latency by replaying block-level I/O traces, which ignores system-level informations and make storage subsystems disconnected from the host, the proposed \texttt{SimpleSSD} can report detailed information from low-level memory to each firmware module to host-side CPU performance with an execution of entire host applications.
%\section{Background}
%\label{sec:background}
%\input{background-ATC}

\section{SimpleSSD}
\label{sec:highlevelview}

Figure \ref{fig:HIL} shows a high-level view of \texttt{SimpleSSD} and explains how our simulator processes the incoming I/O requests. A request is first taken by the host interface layer (HIL), and the corresponding target address is translated by the flash translation layer (FTL). The parallelism allocation layer (PAL) then services the request by abstracting the physical layout of interconnection buses and flash dies. The completion of an I/O request is reported from PAL to the host-side controller via HIL. 
%In this section, we will describe the functional design of each module, aforementioned.  

\subsection{Fully-Functional Firmware Simulation}

\noindent \textbf{Host interface layer.} 
In \texttt{SimpleSSD}, HIL first receives an incoming request from the disk controller of gem5 and enqueues the request in a device-level queue. During this phase, it parses the host-side information and translates it a logical block address (LBA), request type, number of sectors, and a host's system time information (e.g., tick). HIL then forwards this translated information to the underlying FTL through communication APIs, \texttt{ReadTransaction()} and \texttt{WriteTransaction()}. Since there are many different types of simulation models for a full system (e.g., discrete event-driven, activity-driven, and continuous), HIL exposes all request completions through a latency map table, which includes the finish time (i.e., finishTick) along with each requested address. 
%One of the things that HIL needs to take into account is the different simulation domains between a full system simulator and our SSD simulation model. For example, while the execution of an instruction can be stalled in cases where it accesses the underlying SSD (because of a cache miss or system call), the actual full system simulation should not be halted due to I/O service model. 
%To address this issue, HIL employs a latency map table that includes the completion time for each requested address. 
Once the latency for each request is updated by the underlying simulation modules, HIL updates the table with the completion time, and the full-system simulator (e.g., gem5) retrieves it in an asynchronous fashion. While the current queue implementation of HIL is first-come-first-served, system and computer architects can insert their buffer cache, I/O reordering logic, or scheduler into HIL \cite{jung2014hios, O3, jung2014sprinkler, Elyasi}.

\noindent \textbf{Flash translation layer.}
%While a host can send an I/O request whose size varies based on the decision of a full system simulator's file system or traces 
%all actual services at flash level should be performed by page granularity. 
The I/O sizes requested by a host application vary and can be even larger than the page size that a single flash die could accommodate.  
Therefore, in this work, FTL separates the request forwarded by HIL into multiple \emph{sub-requests}, each indicated by a logical page number (LPN). If it is a read, FTL directly translates the sub-requests' LPNs to physical page numbers (PPNs) by looking up its own address mapping table. 
Otherwise, FTL allocates new page(s) and updates the table with appropriate block and/or page addresses and other meta-data information. 
In \texttt{SimpleSSD}, this address translation mechanism is implemented in a functional API, called \texttt{FTLmapping()}. 
The translated or allocated PPNs are then issued into the underlying module's queue by calling \texttt{SendRequest()}, and FTL repeats this process until there is no waiting sub-request. 
When there is no available page for a write, FTL performs garbage collection (GC) to reclaim a set of new pages in flash block(s). 
At the beginning of GC, it selects the victim blocks and free block(s) to allocate as a new block, which can be determined by a wear-leveling algorithm. After this selection, FTL reads the data from all valid pages of the victim blocks, writes them into the new block,  and updates the address table for the reclaimed blocks. Note that the additional read and write operations imposed by GC(s) are treated just like other sub-requests from PAL viewpoint, but the latency associated with all the internal I/O requests is aggregated and exhibits long tail from FTL and HIL perspectives.
%as far as FTL is concerned in \texttt{SimpleSSD}. 
In this work, we consider a simple GC algorithm (cf. greedy), which selects a victim block with the maximum number of invalid pages. The number of free blocks and GC threshold can be reconfigured based on user inputs.
Besides, the wear-leveling algorithm we implemented always allocates new block(s) by considering the minimum erase count among the free blocks in a reserved pool. Users can replace these algorithms with advanced mechanisms \cite{Shahidi, bgc} by updating the \texttt{GarbageCollection()} and \texttt{WearLeveling()}. 

%\cite{bux2010performance}

\subsection{Hardware Simulation for Scalable SSD Parallelism}

\noindent \textbf{Parallelism abstraction layer.}
%From a functionality perspective, the key role of FTL is to translate the addresses, while not managing different levels of parallelism or flash transactions. 
In this work, we introduce PAL underneath FTL and decouple SSD parallelism from other flash firmware modules for improved simulation efficiency and a better research-wise structure. PAL basically stripes all incoming requests across different channels, packages and dies, based on user configurations, which is similar to the striping method employed by RAID. At the beginning, PAL dequeues the requests issued by FTL and disassembles the target page address by being aware of the underlying hardware configuration (e.g., numbers of channels, flash packages, and dies). This is implemented with \texttt{PPNdisassemble()}.
Based on the disassembled information, PAL simulates SSD internal state and schedules the flash transaction at a finer granularity to capture the memory-specific latency, idle time, and even scheduling penalties imposed by resource contentions. In other words, the latency of a sub-request can be dynamically simulated in \texttt{SimpleSSD} by considering not only the hardware resource availability but also the storage media configuration. After processing the I/O request, PAL returns the simulated latency for each sub-request to FTL. FTL then collects and reevaluates them to generate an appropriate latency for the I/O request that possesses such sub-requests. By being aware of the states of the underlying hardware, users can explore new parallelism strategies and schedulers. The order for sub-request striping or management of flash transactions can be determined by modifying \texttt{PPNdisassemble()} and  \texttt{TimelineScheduling()}, respectively.

\begin{figure}
\centering
%\subfloat[ISPP.]{\label{figs:readlatency}\rotatebox{0}{\includegraphics[width=0.24\linewidth]{figs/ispp-vth}}}
%\hspace{1pt}
%\subfloat[TLC Programming.]{\label{figs:readlatency}\rotatebox{0}{\includegraphics[width=0.24\linewidth]{figs/ispp-dist}}}
%\hspace{1pt}
\subfloat[Read latency.]{\label{figs:readlatency}\rotatebox{0}{\includegraphics[width=0.49\linewidth]{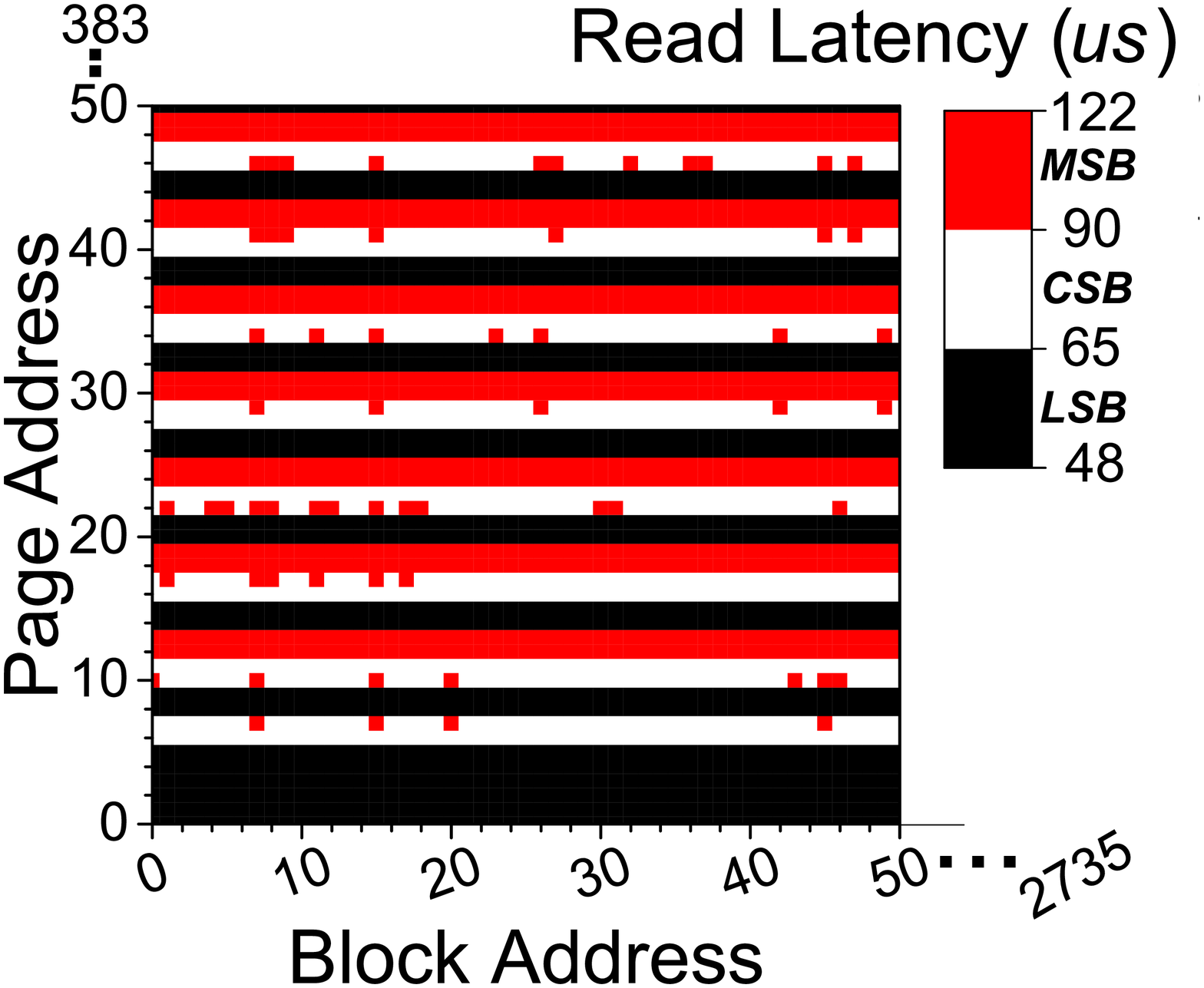}}}
\hspace{1pt}
\subfloat[Write latency.]{\label{figs:writelatency}\rotatebox{0}{\includegraphics[width=0.49\linewidth]{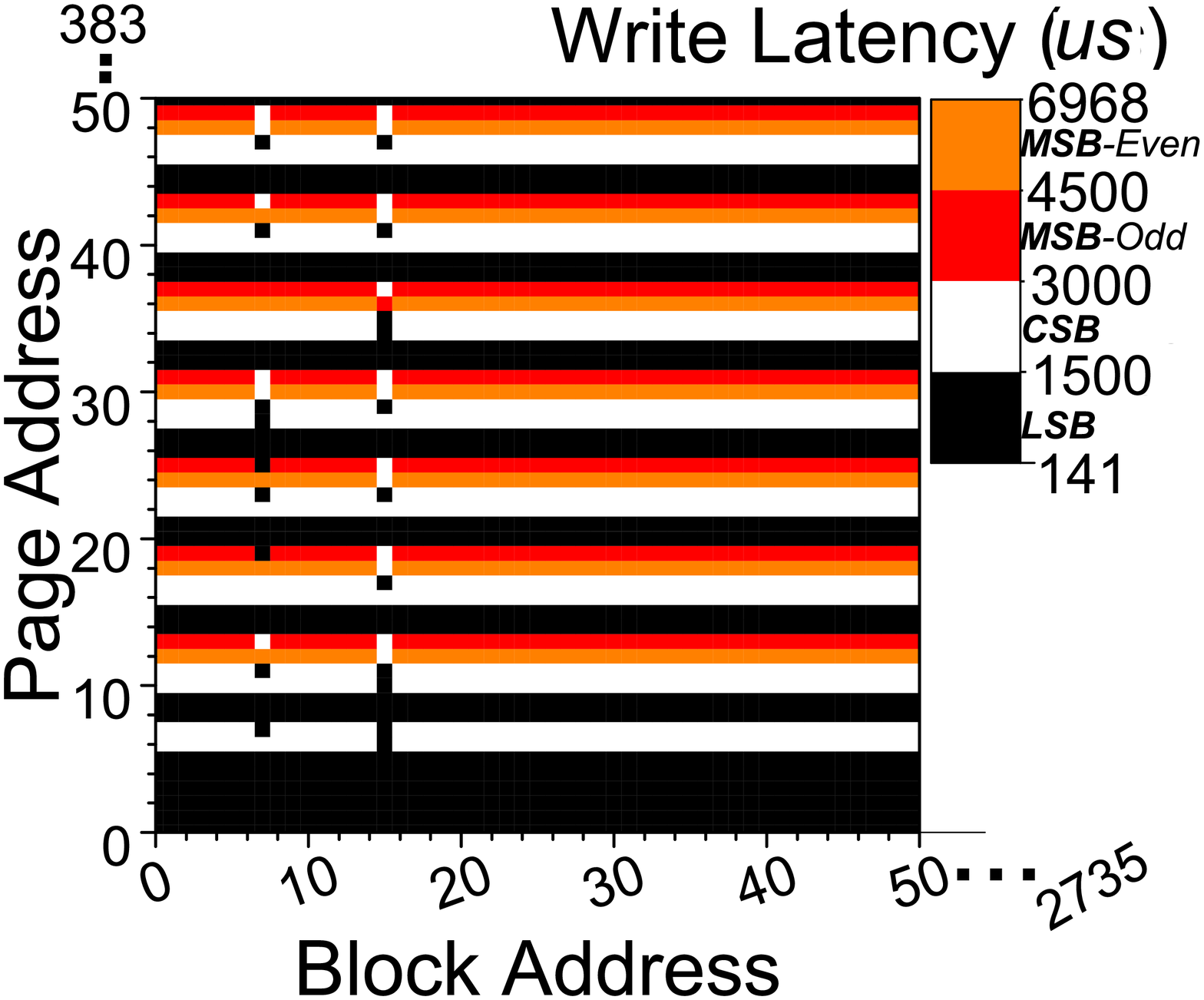}}}
\caption{\label{fig:latency}Analysis for flash intrinsic latency variation.}
\end{figure}

\begin{table}
\centering
\resizebox{\textwidth}{!}{
\caption{Configurations for system simulations.}
\label{tab:sys_config}
\begin{tabular}{|c|c|c|c|c|c|l|}
\hline
\textbf{Channel} & \textbf{Package} & \textbf{Die} & \textbf{Plane} & \textbf{Block} & \textbf{CPU} & ARMv7, 1 core, 1GHz \\ \hline
8 & 8 & 4 & 2 & 1024 & \multirow{2}{*}{\textbf{Cache}} & L1I\$/L1D\$, 64KB, 4-way \\ \cline{1-5} \cline{7-7} 
\textbf{DMA(MHz)} & \textbf{Pages/blk} & \textbf{Page} & \textbf{OP} & \textbf{GC} &  & L2\$, 512KB, 8-way \\ \hline
400 & 256 & 8KB & 0.2 & 0.05 & \textbf{DRAM} & DDR3, 1 channel, 2 rank \\ \hline
\end{tabular}}
\end{table}

\begin{table*}[]
\centering
\resizebox{\textwidth}{!}{
\caption{Important characteristics of the tested benchmarks.}
\label{tab:workload}
\begin{tabular}{|l|c|c|c|c|c|c|c|c|c|c|c|c|c|}
\hline
\textbf{Workloads} & \textbf{apache1} & \textbf{fileserver1} & \textbf{fileserver2} & \textbf{fileserver3} & \textbf{fileserver4} &  \textbf{varmail1} & \textbf{varmail2} & \textbf{varmail3} & \textbf{varmail4} & \textbf{webserver1} & \textbf{webserver2} & \textbf{iozone} & \textbf{mmap} \\ \hline
\textbf{Storage/K Instruction} & 26 & 82 & 127 & 86 & 126 & 8 & 6 & 7 & 6 & 5 & 4 & 57 & 109 \\ \hline
\textbf{SSD Read Ratio (\%)} & 99 & 5.5 & 2.2 & 6.1 & 2.3 & 60 & 74 & 60 & 73 & 99 & 99 & 4 & 51 \\ \hline
\textbf{Max Instructions (B)} & 5 & 18 & 5 & 17 & 5 & 3 & 3 & 3 & 3 & 3 & 3 & 4 & 3 \\ \hline
\end{tabular}}
\end{table*}

%\subsection{Enabling Full System Simulation}

\noindent \textbf{Latency variation mapping.}
To make the storage denser with the same number of transistors, flash can store multiple states into a single storage cell. 
For example, \emph{triple-level cell} (TLC) flash stores eight different states into a target storage core. Each state is represented by different voltage thresholds ($V_{th}$). Because a TLC core can maintain 3-bit data, the TLC technology can drastically increase the storage capacity of an SSD. However, the materials of the TLC storage core are not fundamentally different from that of a \emph{single-level cell} (SLC) or \emph{multiple-level cell} (MLC), which can represent 1-bit or 2-bit data per cell, respectively. Instead, the flash logic of TLC (and MLC) writes (i.e., programs) data into a target in a different manner compared with SLC flash. This is referred to as an \emph{increment step pulse program} (ISPP \cite{suh19953}) and introduces significant latency variation.
To characterize the latency behavior incurred by the ISPP, we built an FPGA-based controller by using Xilinx Spartan-6 and tested SLC, MLC, and TLC NAND flash devices. Figures \ref{figs:readlatency} and \ref{figs:writelatency} illustrate the latency variation observed for writes and reads on TLC 25 nm flash technology \cite{TLC-NAND}, respectively; we provide only the TLC results owing to the page limit, but other flash technologies also exhibited the same latency trend that we observed for TLC. The evaluation data were measured for every single block and page. For writes, the latency of the \emph{most significant bit} (MSB) pages was longer than those of the \emph{center significant bit} (CSB) and \emph{least significant bit} (LSB) pages by approximately 1.3 and 8 times, respectively. 
%The reason why there exists the significant latency disparity among LSB/CSB/MSB is that TLC flash preforms many verify-and-program steps (of ISPP) to generate an accurate voltage distribution on the target. However, each state of TLC flash has a very tight $V_{th}$ margin as considering that it puts eight states into a single cell, and therefore, such verify-and-program process introduce a number of reads and programs to write appropriate data. 
The reads on TLC flash also exhibited similar latency variation characteristics. Specifically, the read latency of MSB pages is longer than that of CSB pages and LSB pages by 37\% and 84\%, on average, respectively. 
%This is because flash control logic (within a chip) requires to check the target state by shifting and applying the reference voltages till to find out the target. Thus, the number of checks on MSB pages is greater than that of CSB pages and LSB pages. 
Since the latencies between different pages exhibit a notable difference, this can have a great impact on parallelism and hardware modeling. We observed that the first five pages within a block always exhibited LSB page performance, and the latency of the next three pages (i.e., after the first five) was the same as that of the CSB pages. 
These eight pages, referred to as \emph{meta pages}, are usually used for storing the metadata of flash firmware, such as mapping information associated with the block. 
The latency for all remaining pages can be mapped with the following simple function: $f(addr) = (addr - n_{meta}) / n_{plane} \bmod n_{state}$ where $addr$, $n_{meta}$, $n_{state}$ and $n_{plane}$ are the input address, number of meta pages, number of states per cell and number of planes within a flash die, respectively. If $f(addr)$ is 0, it is an LSB page. If $f(addr)$ is 1, it is a CSB page. Otherwise, the address indicates an MSB page.

%\section{Model Implementation}
%\label{sec:implementation}
%\input{implementation}

\begin{figure}
\centering
%\def\subfigcapskip{0pt}
%\vspace{-15pt}
\subfloat[Writes.]{\label{figs:BW-verify-w}\rotatebox{0}{\includegraphics[width=0.48\linewidth]{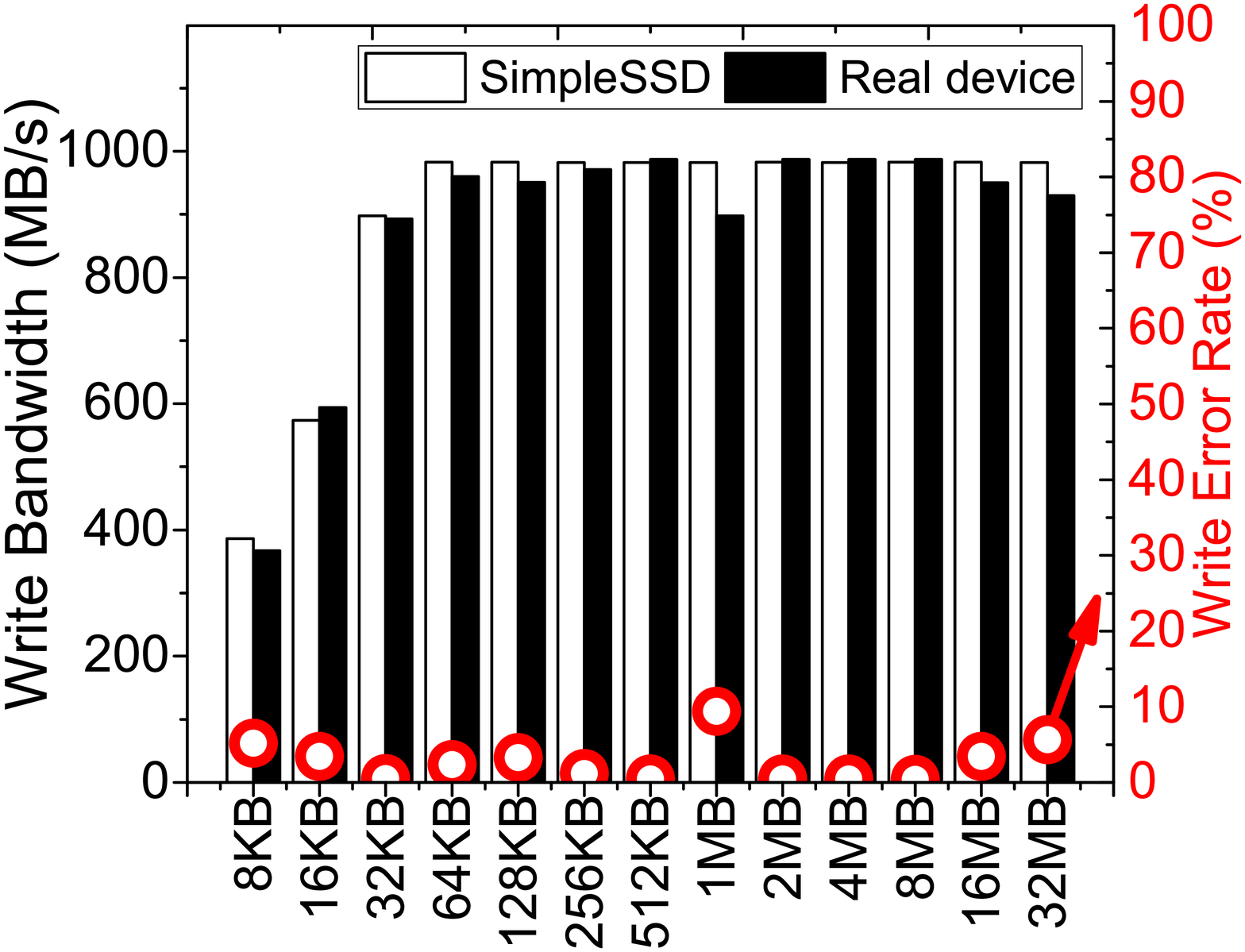}}}
\hspace{2pt}
\subfloat[Reads.]{\label{figs:BW-verify-r}\rotatebox{0}{\includegraphics[width=0.48\linewidth]{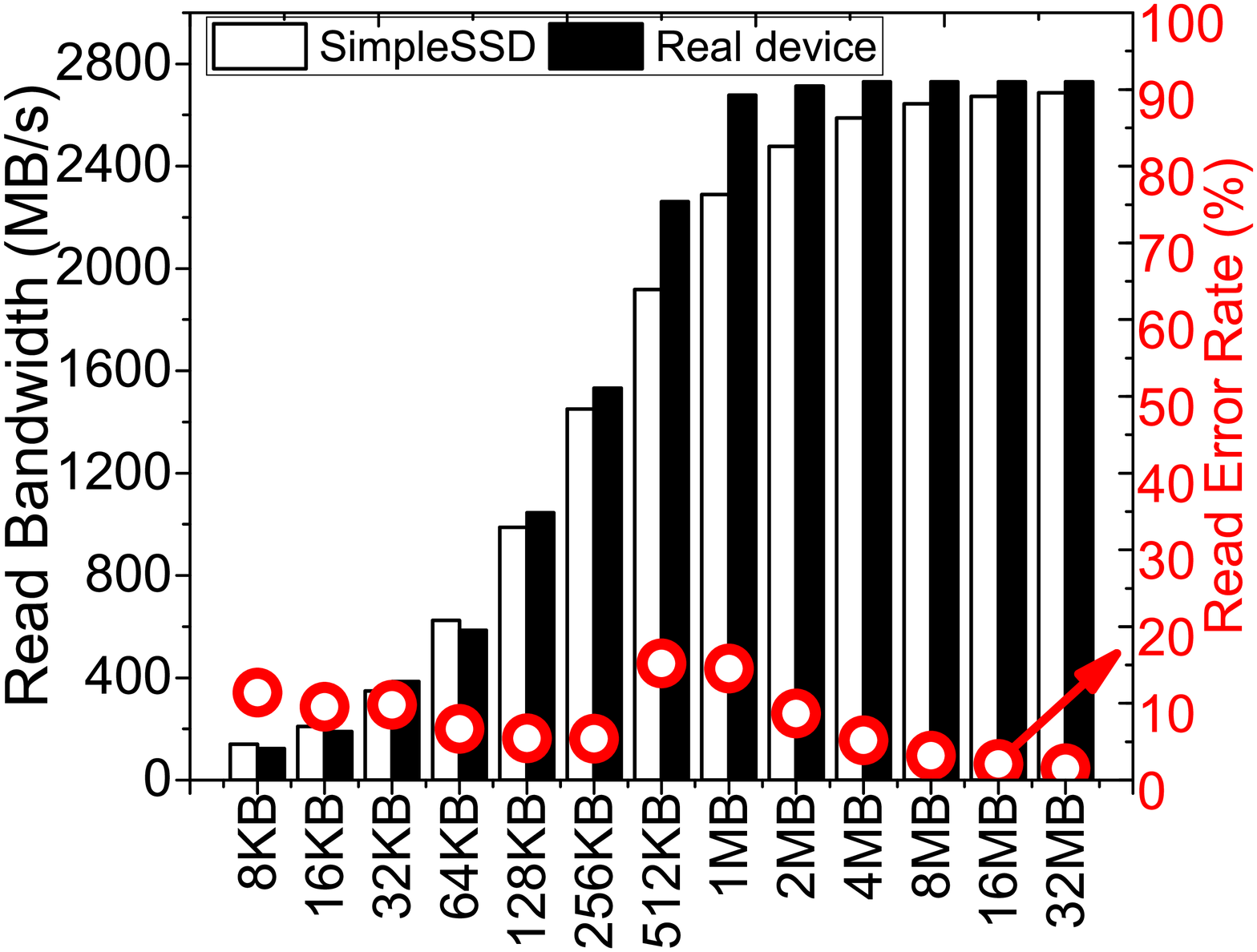}}}
\caption{\label{fig:BW-verify}Set of evaluations for performance validation. }
\end{figure}

\section{Evaluation}
\label{sec:evaluation}

\noindent \textbf{System devices and software configurations.}
We configure a host that employs an eight-bank DDR3-1600 DRAM and 1GHz CPU (ARM). The underlying storage is configured as an eight-channel high performance SSD device. Each channel connects eight packages, each with four TLC flash dies. 
FTL of this baseline is configured with a set-associative mapping algorithm, which associates eight log blocks with a single physical block. FTL has 20\% over-provisioning (OP) space, and its GC threshold is set to 5\%. The detailed information for system configurations, including CPU, SSD and flash, are given by Table \ref{tab:sys_config}. Lastly, we simulate SSDs with Linux 3.13.0 and EXT2 file system driver.

\begin{figure*}
\centering
%\def\subfigcapskip{0pt}
%\vspace{-15pt}
\subfloat[IPC.]{\label{figs:NAND_IPC}\rotatebox{0}{\includegraphics[width=0.24\linewidth]{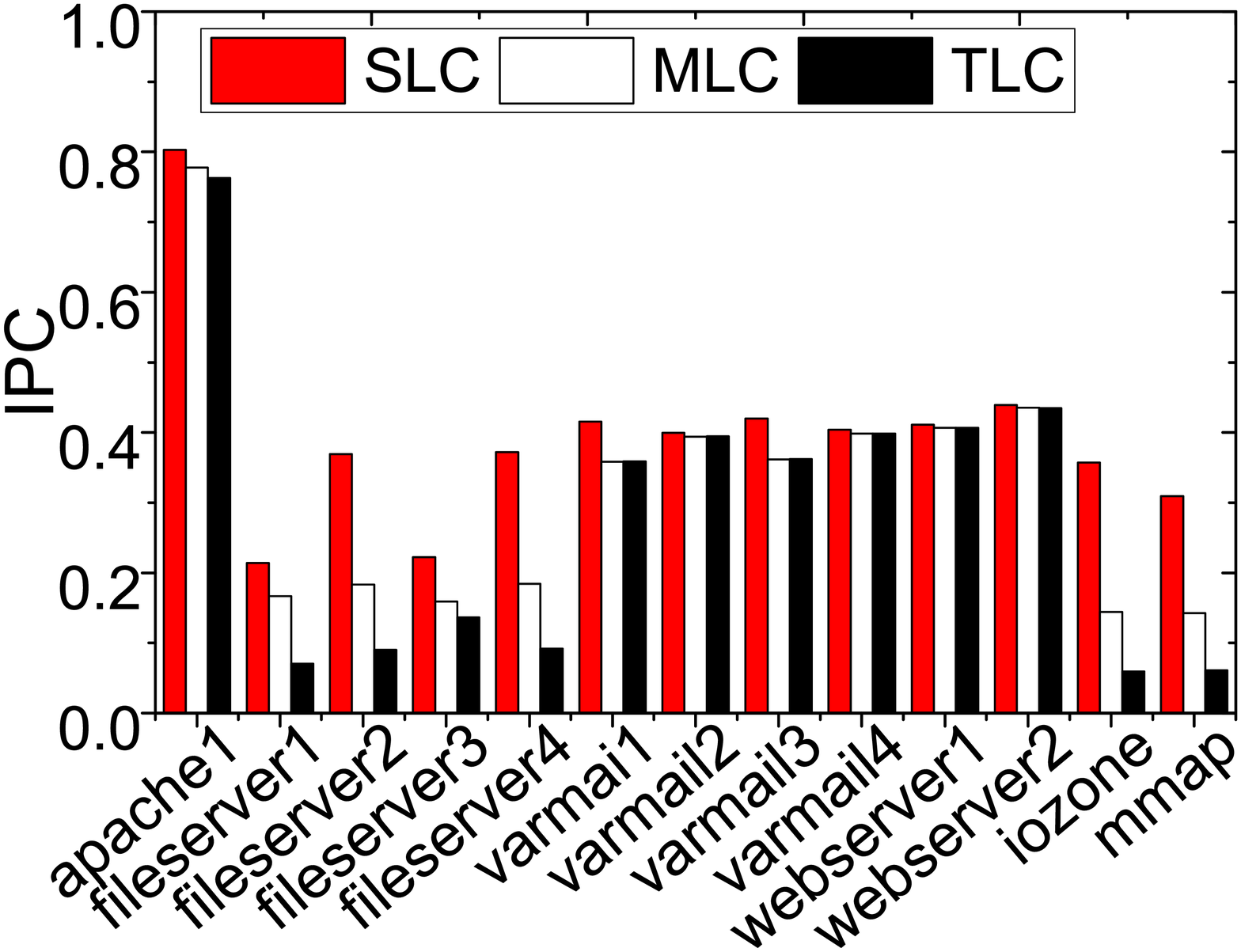}}}
\hspace{1pt}
\subfloat[Data access breakdown.]{\label{figs:NAND_volumebrk1}\rotatebox{0}{\includegraphics[width=0.24\linewidth]{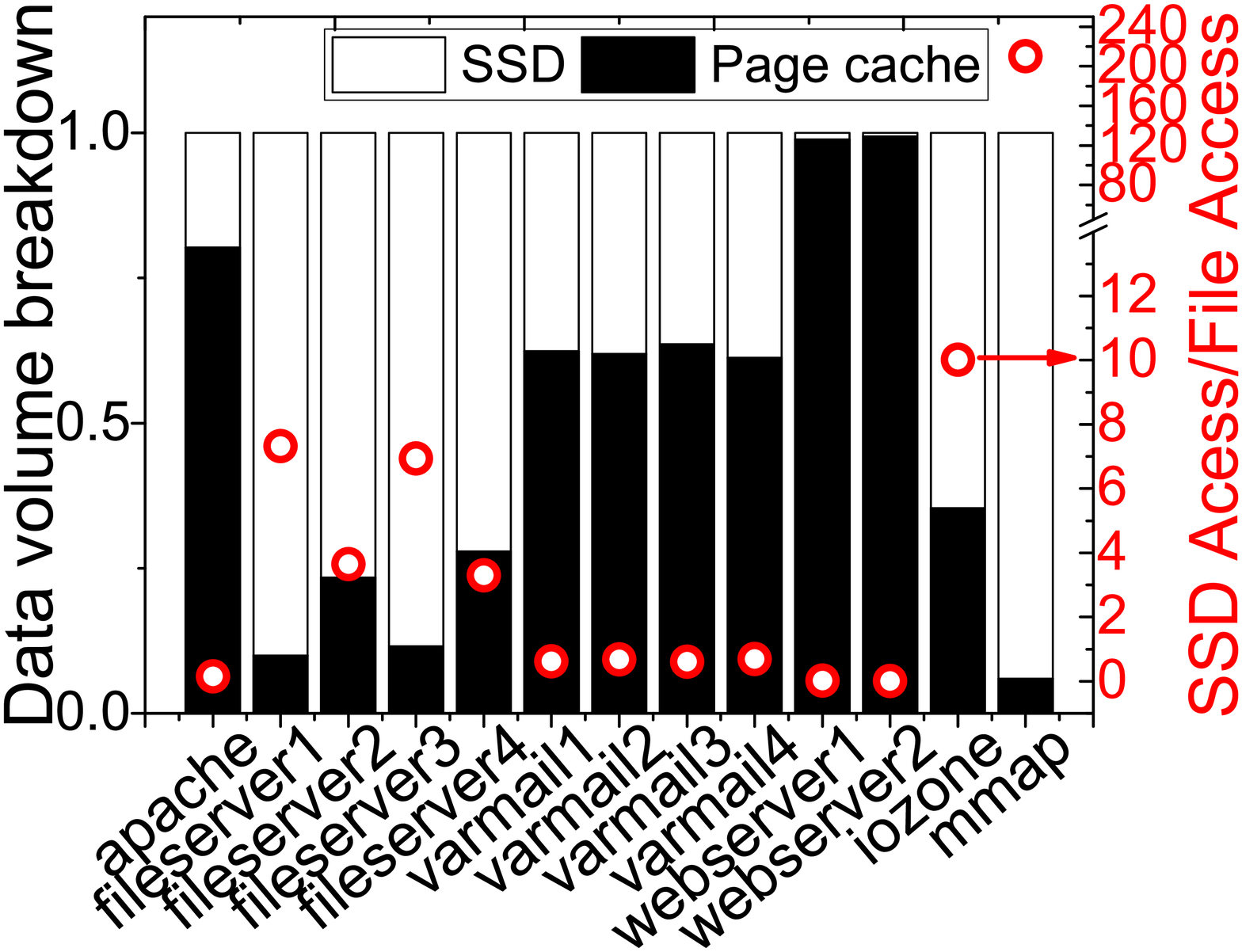}}}
\hspace{1pt}
\subfloat[System breakdown.]{\label{figs:NAND_sysbrk1}\rotatebox{0}{\includegraphics[width=0.24\linewidth]{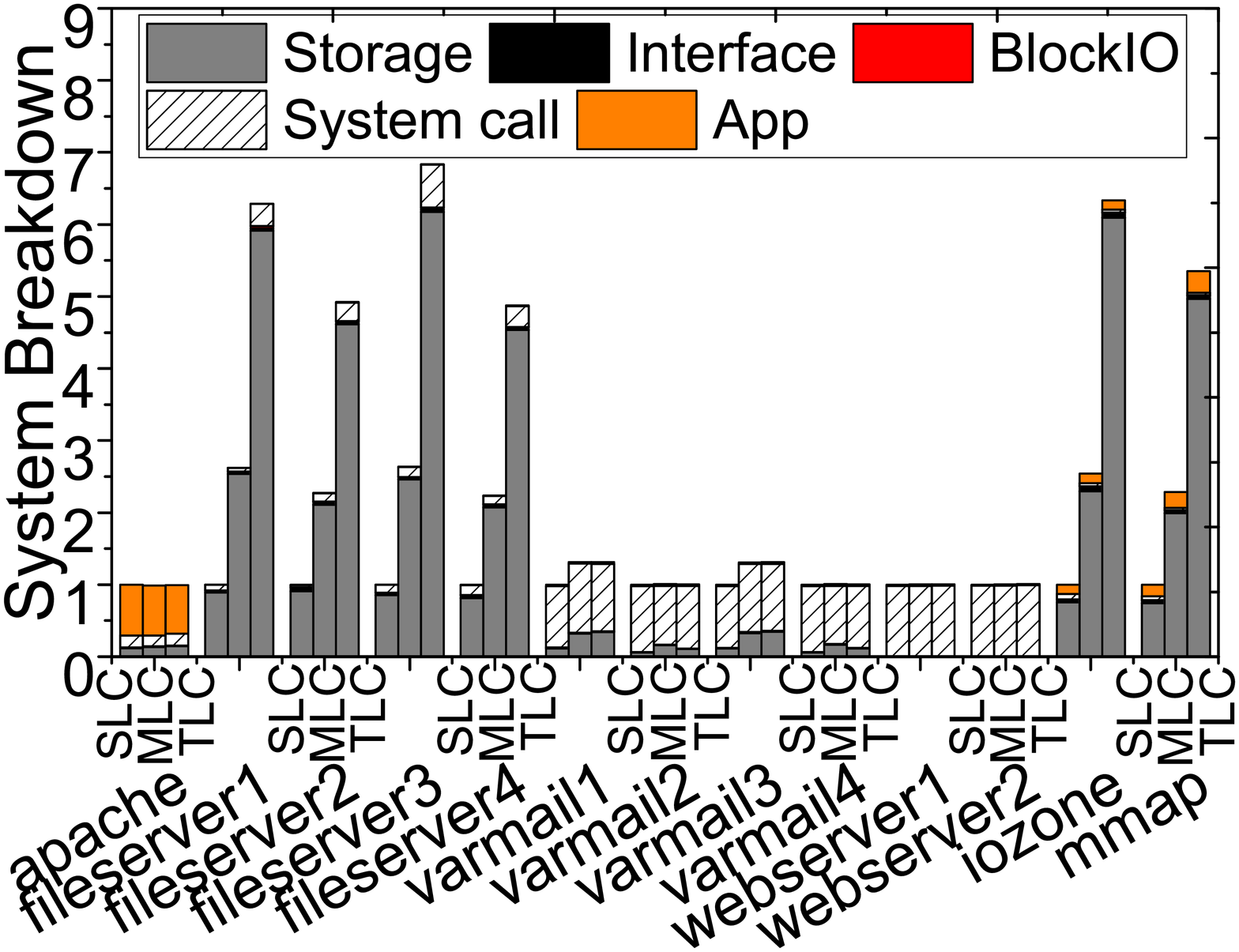}}}
\hspace{1pt}
\subfloat[SSD latency breakdown.]{\label{figs:NAND_SSDlat}\rotatebox{0}{\includegraphics[width=0.24\linewidth]{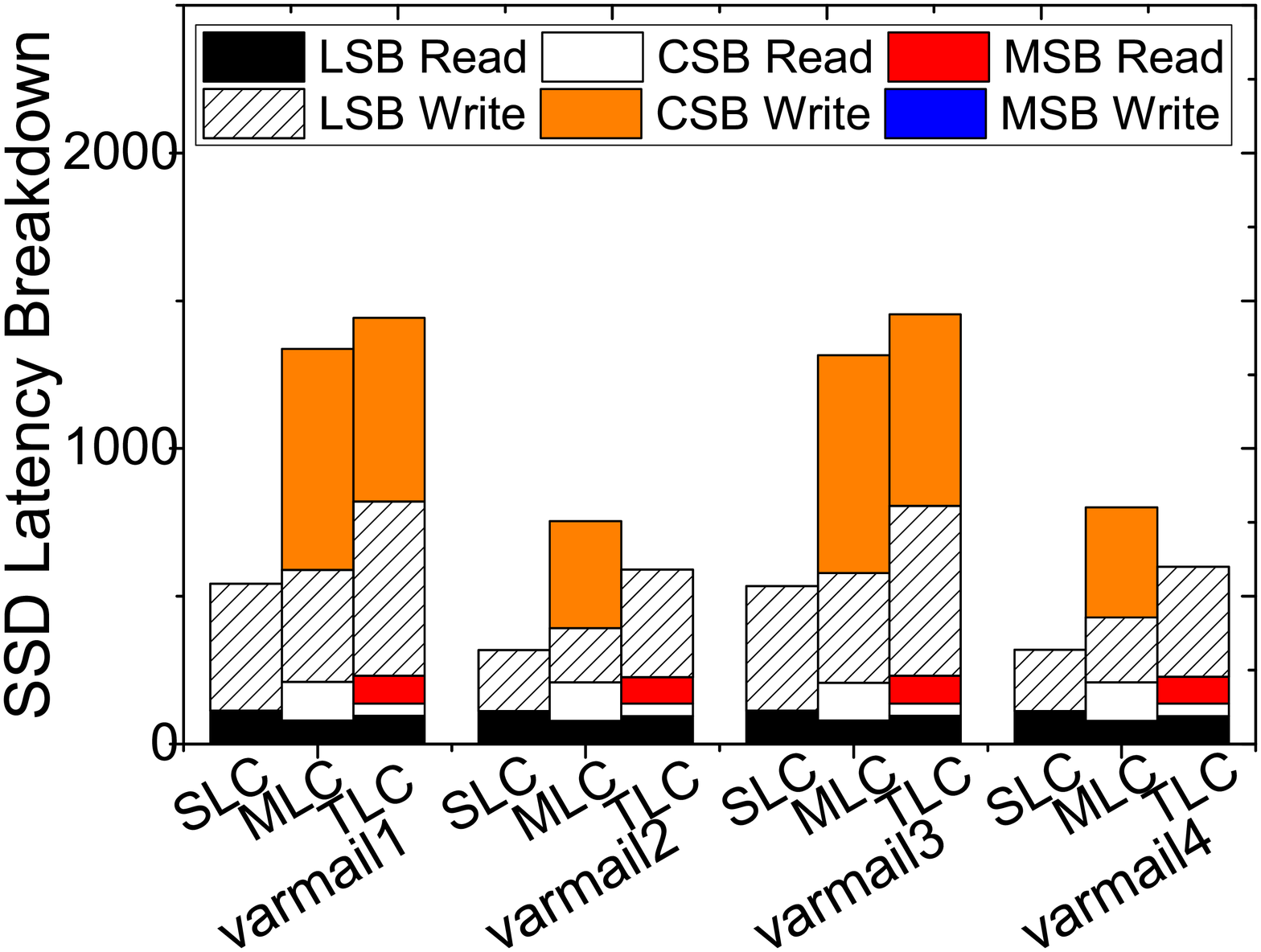}}}
\caption{System-level performance analysis with three different non-volatile memory technologies. \label{fig:IPC-brk}}
\end{figure*}

\noindent \textbf{Workloads.} In this evaluation, we use 13 different workloads. 
% with real executions of apache \cite{}, iozone filesystem \cite{}, mmap \cite{} and filebench \cite{} benchmark suites. 
Specifically, \emph{Apache}Bench \cite{APACHE} is used to measure the performance of an HTTP web server, where a specified URL is processed by generating heavy storage reads for the corresponding HTTP file(s). \emph{Filebench} \cite{mcdougall2005filebench} includes several storage-centric workloads; each creates, writes and reads a few thousand files. In addition to these basic file I/Os, \emph{fileserver} appends data and performs several file-sync operations with multiple threads, whereas \emph{varmail} and \emph{webserver} repeatedly read 1,000 small-sized files and write logs. Compared with \emph{webserver}, \emph{varmail} has extra I/O operations related to file deletion and creation. Finally, \emph{Iozone} \cite{norcott2003iozone} evaluates a file system with a given automatic mode, and \emph{mmap} \cite{mmap} keeps reading and writing many files over POSIX library's APIs. Table \ref{tab:workload} lists the important characteristics of these workloads.

\subsection{Performance Validation.}
%We validate the basline SSD simulation with an NVMe SSD device \cite{NVMeSSD} coming from Intel. Like all other SSD vendors, this Intel NVMe 750 device does not offer official hardware configuration in detail at all, but we conjuncture that the number of channels and packages can be reasonablly similar to our baseline \cite{NVMeSSD}. Since the storage performance can also vary based on buffers and file systems in storage stack, we disable write cache using ``O\_DIRECT" option in this section. 

We compare the performance of \texttt{SimpleSSD} simulations in standalone mode with that of a real device (Intel 750).  
Specifically, we use multiple storage traces of ATTO \cite{ATTO} to analyze the disk-level characteristics in detail. Basic read and write tests were performed with varying I/O request sizes. Figure \ref{fig:BW-verify} shows the results.
For all requests ranging in size from 8 KB to 32 MB, the percentage difference (i.e., error rate) between the results of \texttt{SimpleSSD} and Intel 750 is 2.7\% on average, and the performance trends are similar. When the request size is increased, the bandwidth of both drives quickly increased and saturates at the 64KB. On the other hand, the percentage difference of the reads is 7.1\% on average. While the performance trends of the two devices are similar, the \texttt{SimpleSSD} performance increases more gradually than that of the real device; this makes the read error rate slightly higher than the write error rate. We conjecture that the real device has  vendor-specific optimization, such as read-ahead or caching. Note that the current version of \texttt{SimpleSSD} has no specific buffer caching algorithm or acceleration model, which can introduce a greater performance disparity (compared to Intel 750) for small-sized I/O request tests.  
In addition to these microbenchmark tests, we also validate \texttt{SimpleSSD} by comparing its performance with that of a real device when executing 14 real storage workloads \cite{verma2010srcmap,kavalanekar2008characterization}, which includes real storage access patterns of a web server, database, and enterprise cluster. We observed that the performance trend of \texttt{SimpleSSD} with these workloads is similar to that of the real device. More practically, for these real workload evaluations, the difference between them is 9\% on average.

\subsection{SSD-Enabled Full System Evaluation}
\noindent \textbf{Overall CPU performance.} Figure \ref{figs:NAND_IPC} shows the CPU performance (IPC) of hosts that employ different flash technologies (i.e., SLC/MLC/TLC) as their storage subsystems. All IPCs are normalized to those of the SLC version. 
As expected, the SLC-equipped system has better IPC than the MLC- and TLC-equipped systems by averages of 44\% and 141\%, respectively. 
%As expected, the SLC-equipped system has average of 44\% and 141\% better IPCs, compared with MLC- and TLC-equipped systems, respectively. 
Interestingly, \emph{apache} and \emph{webserver} show small or almost no performance benefit over SLC. As shown in Figure \ref{figs:NAND_volumebrk1}, even though these servers read many files, most of them are served from VFS's page cache. In contrast, \emph{fileserver}, \emph{iozone} and \emph{mmap} have poor locality regarding the target (i.e., they touch once and never refer again), and have many fsync and/or flush operations, which make the page cache inefficient. A total of the 19\% of I/O accesses is served by the page cache, on average. 
Even though \emph{varmail} also exhibit many reads like \emph{webserver}, it has slightly different performance characteristics. We explain the reason shortly.

\begin{figure}
\centering
\includegraphics[width=1\linewidth]{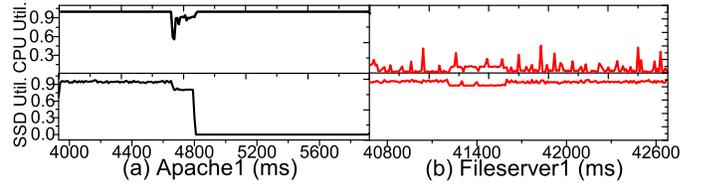}
\caption{Time series analysis.}
\label{fig:NAND_tot_apache_fileserv}
\end{figure}

\noindent \textbf{Storage stack analysis.}
Figure \ref{figs:NAND_sysbrk1} decomposes the execution time spent for each component. It excludes overlaps of time with the latency consumed by the underlying component. For a better comparison, all MLC and TLC values are normalized to SLC ones. As expected, file-intensive benchmarks including \emph{fileserver}, \emph{iozone} and \emph{mmap}, spend the most time accessing the underlying storage. Thus, the SLC-equipped system performs better than the MLC- and TLC-equipped systems by around 2.5x and 5.8x, respectively. However, \emph{apache} shows a completely different performance behavior than \emph{fileserver}. Specifically, it consumes more CPU cycles at the user application level (68\% of the total time) rather than storage accesses. This is because most of the cycles consumed by a block layer and system call overlap with those of underlying storage services, while processing the HTTP service keeps the entire CPU busy. 
For better understanding, we analyze the time series of CPU utilization and SSD utilization, which are measured at the end of benchmark executions for 2s. Compared to \emph{fileserver1}, which utilizes the CPU 11\% of the time on average while utilizing the SSD almost 100\% of the time, \emph{apache} activates CPU constantly. It has many overlaps with the SSD activities. Even after the SSD completes all read services, \emph{apache} continues to process their data, which exhibit a high IPC. %We will analyze the storage stack characteristics of \emph{varmail} below.

\noindent \textbf{Device analysis.} Figure \ref{figs:NAND_SSDlat} shows the page-level latency breakdown for four \emph{varmail} workloads. Interestingly, the write patterns of \emph{varmail2} and \emph{varmail4} have no address associated with CSB and MSB pages. Because all of the writes are served from the LSB pages, the TLC-based SSD has 34\% and 32\% shorter latencies on average, respectively, than the MLC-based SSD. However, these performance benefits are not directly reflected in the IPC, as shown in Figure \ref{figs:NAND_IPC}. This is because, as shown in  Figure \ref{figs:NAND_sysbrk1}, most of the time spent by \emph{varmail} is consumed by system calls, which are primarily related to handling the page cache. This time consumed by the system calls, which does not overlap with the underlying device operations, accounts for more than 90\% of the overhead for all executions.

\subsection{Related and Future Work}
There are very few SSD simulators in literature that are publically available for download \cite{prabhakaran2009ssd, flashsim, jung2012nandflashsim, hu2011performance}. Even with these simulators, constraints prevent design space exploration for emerging memory/storage hierarchies. 
First, the hardware organization of existing simulators \cite{prabhakaran2009ssd, flashsim} is unfortunately overly-simplified and far from capturing the critical features of high-performance contemporary SSD architectures. There is neither a specific flash microarchitecture nor an internal parallelism model. In addition, these simulators cannot fully reflect the important functionalities of the underlying flash firmware, which also have a great impact on system performance. The simulators have no FTL \cite{hu2011performance, jung2012nandflashsim} or an ideal FTL \cite{prabhakaran2009ssd}. Note that none of these existing SSD simulators can be directly used for full system simulations. %In addition, it is also difficult to integrate all flash firmware and hardware characteristics into a full system simulator. 

In contrast, our \texttt{SimpleSSD} not only models contemporary SSDs by employing a complete storage stack and detailed hardware parallelism but also enables system-level simulation by considering different flash memory technologies. Thus it enables researchers to study diverse system performance characteristics from a holistic viewpoint. 

\noindent \textbf{Future work.} Computer Architecture and Memory Systems Laboratory (CAMEL) is extending the current simulation framework by implementing new features such as PCIe-enabled system/IO crossbars, message-signaled interrupts, internal DRAM models, NVMe interfaces and memory power models.

%\begin{figure*}
%\centering
%%\def\subfigcapskip{0pt}
%%\vspace{-15pt}
%\subfloat[Time series analysis of Apache.]{\label{figs:NAND_tot_apache4}\rotatebox{0}{\includegraphics[width=0.49\linewidth]{figs/NAND_tot_apache4}}}
%\hspace{1pt}
%\subfloat[Time series analysis of Fileserver1.]{\label{figs:NAND_tot_fileserver4}\rotatebox{0}{\includegraphics[width=0.49\linewidth]{figs/NAND_tot_fileserver4}}}
%\vspace{-10pt}
%\caption{\label{fig:Utilization} \vspace{-15pt}}
%\end{figure*}

%\begin{figure}
%\centering
%%\def\subfigcapskip{0pt}
%%\vspace{-15pt}
%\subfloat[CPU utilization.]{\label{figs:NAND_CPUutil}\rotatebox{0}{\includegraphics[width=0.49\linewidth]{figs/NAND_CPUutil}}}
%\hspace{1pt}
%\subfloat[Storage utilization.]{\label{figs:NAND_storageutil}\rotatebox{0}{\includegraphics[width=0.49\linewidth]{figs/NAND_storageutil}}}
%\vspace{-10pt}
%\caption{\label{fig:Utilization} \vspace{-5pt}}
%\end{figure}

%\section{Related Work}
%\label{sec:related}
%\input{related}

\section{Acknowledgement}
This research is mainly supported by NRF 2016R1C1B2015312. This work is also supported in part by IITP-2017-2017-0-01015, NRF-2015M3C4A7065645, DOE DE-AC02-05CH 11231, and MemRay grant (2015-11-1731). Dr. Kim is supported in part by NSF 1640196 and SRC/NRC NERC 2016-NE-2697-A. Dr. Kandemir is supported in part by NSF grants 1439021, 1439057, 1409095, 1626251, 1629915, 1629129 and 1526750. Myoungsoo Jung is the corresponding author. 

\section{Conclusion}
\label{sec:conclusion}
We proposed a high-fidelity SSD simulator that builds a complete storage stack from scratch and models all detailed characteristics of SSD internal hardware and software. This simulator can be integrated into publicly-available full system simulators.

\ifCLASSOPTIONcaptionsoff
  \newpage
\fi

%%%%%%%%% -- BIB STYLE AND FILE -- %%%%%%%%
\bibliographystyle{abbrv}
\bibliography{SimpleSSD}
\end{document}